\documentclass[twocolumn,amsmath,amssymb]{revtex4-1}
\usepackage{txfonts}

\usepackage{bm}
\usepackage[dvipdfmx]{graphicx}
\usepackage{color}
\usepackage{braket}

\usepackage{mathdots}

\usepackage[pdftex]{hyperref}
\hypersetup{
	colorlinks=true
}

\renewcommand{\(}     {\left(}
\renewcommand{\)}     {\right)}

\renewcommand{\_}[1]  {_\textrm{#1}}

\begin{document}

\title{
Robust magnetotransport in disordered ferromagnetic kagome layers\\ with quantum anomalous Hall effect
}
\author{Koji Kobayashi$^1$}
\author{Masaki Takagaki$^1$}
\author{Kentaro Nomura$^{1,2}$}
\affiliation{$^1$Institute for Materials Research, Tohoku University, Sendai 980-8577, Japan}
\affiliation{$^2$Center for Spintronics Research Network, Tohoku University, Sendai 980-8577, Japan}

\begin{abstract}
 The magnetotransport properties of disordered ferromagnetic kagome layers are investigated numerically.
 We show that a large domain-wall magnetoresistance or negative magnetoresistance
can be realized in kagome layered materials (e.g.~Fe$_3$Sn$_2$, Co$_3$Sn$_2$S$_2$, and Mn$_3$Sn),
which show the quantum anomalous Hall effect.
 The kagome layers show a strong magnetic anisotropy and a large magnetoresistance depending on their magnetic texture.
 These domain-wall magnetoresistances are expected to be robust against disorder and observed irrespective of the domain-wall thickness, in contrast to conventional domain-wall magnetoresistance in ferromagnetic metals.
\end{abstract}

\maketitle

\textit{Introduction.}
 Racetrack memory \cite{Parkin08magnetic} has been expected to be a new generation spintronics memory device,
which consists of a ferromagnetic wire with magnetic domains corresponding to ``0'' and ``1''.
 Those domains can be driven by electric current and be read out rapidly without mechanical heads.
 The most prominent advantage of spin memories is that the device does not need electricity to keep information as conventional random access memories.
 That is, the racetrack memory can become a low-power-consumption device.
 However, this ferromagnetic metal spintronics device is not practically realized yet,
in contrast with magnetic tunnel junction devices using the giant magnetoresistance effect 
\cite{Parkin95giant},
which have achieved great success.
 The problem is that a current driven device of ferromagnetic metals suffers from Joule heating.
 In this paper, we propose that we may overcome this difficulty by using the quantum anomalous Hall (QAH) \cite{Haldane88model, Chang13experimental} or Weyl semimetal \cite{Wan11, Burkov11} state of kagome layered materials.
 In those topological states, domain walls can be driven by the electric field \cite{Upadhyaya16domain, Araki16universal, Kurebayashi19theory, Kim19electrically}.
 Therefore a large electric current that causes Joule heating is not necessary.
 We show, via a numerical simulation of transport, that the current is strongly suppressed by domain walls in QAH kagome layers.

 In ferromagnetic metals, the domain-wall magnetoresistance (DWMR) effect originates from the spin mistracking of the conduction electrons \cite{Kent01domain,MaekawaBook}.
 Therefore the DWMR is fragile against a gradual change of magnetization (thick domain walls) or disorder,
and is a weak effect compared with the widely utilized giant magnetoresistance effect \cite{Yavorsky02giant}.
 Nevertheless, the DWMR in magnetic Weyl semimetals \cite{Hirschberger16, Wang16, Jin17ferromagnetic} has been found to become huge in thick domain walls and robust against disorder \cite{Ominato17anisotropic, Kobayashi18helicity}.
 Recently, it was proposed that the QAH state can be realized in kagome layered materials:
 Fe$_3$Sn$_2$ \cite{Ye18massive,Yin18giant}, Co$_3$Sn$_2$S$_2$ \cite{Liu18giant, Muechler18prediction, Yin19negative, Ozawa19two, Liu19magnetic}, and Mn$_3$Sn \cite{Nakatsuji15large, Yang17topological, Ito17anomalous}.
 In order to utilize these materials for novel topological spintronics devices,
we reveal the mechanism and robustness of the DWMR in kagome layers.

 In this paper, we propose a huge DWMR effect
in ferromagnetic kagome layers under disorder.
 We first study the transport property with in-plane or out-of-plane magnetizations.
 Next we show the magnetization angle induced topological phase transition.
 Then we investigate the DWMR for various types of domain wall.
 We show that the DWMR is robust against disorder
and hardly suppressed by thick domain walls.
 These results imply that the DWMR comes from the topological transport in kagome layers.

\begin{figure}[tbp]
 \centering
  \includegraphics[width=1\linewidth]{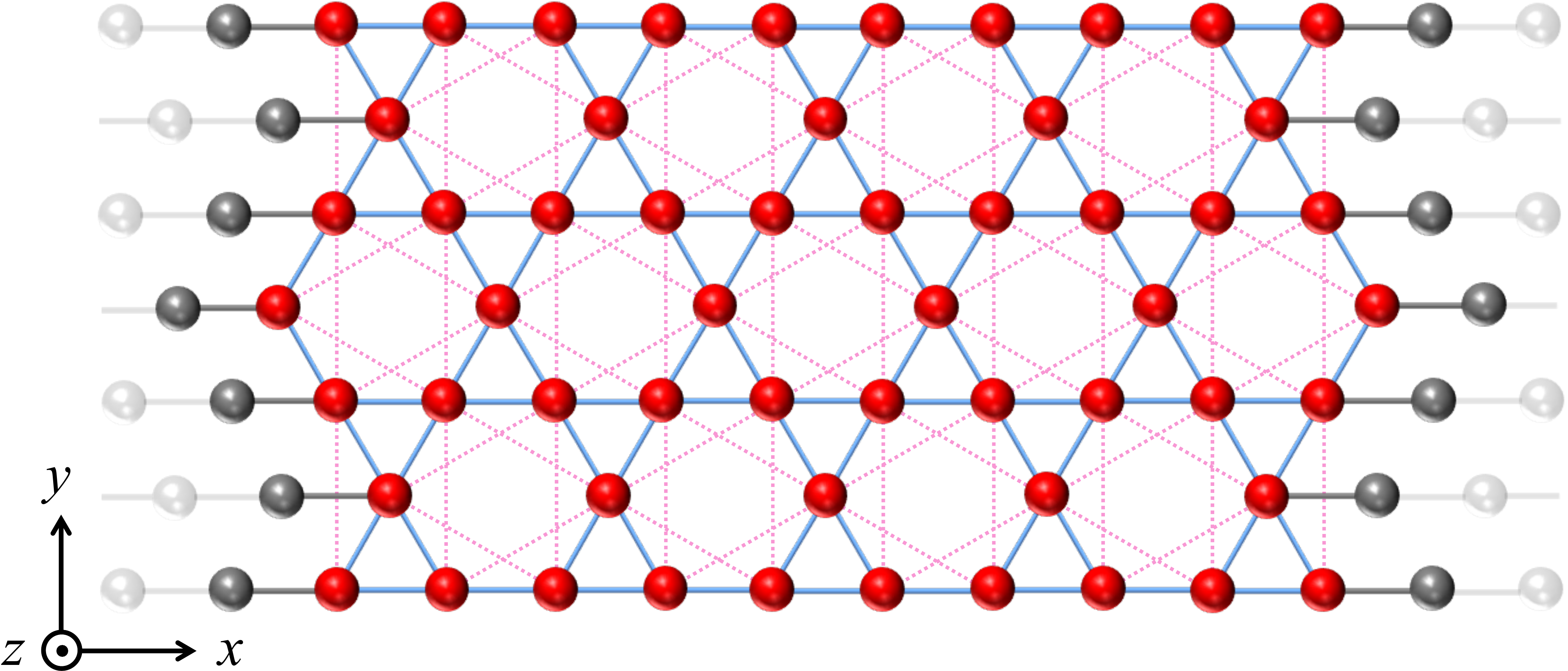}
 \vspace{-3mm}
\caption{%
  Schematic figure of two-terminal geometry of the straight-edged kagome ribbon (red sites).
  One-dimensional ideal leads (gray sites) are attached on both zigzag ends.
  The cyan and pink bonds represent the nearest- and next-nearest-neighbor hopping, respectively.
}
\label{fig:system}
\end{figure}


\textit{Model.}
 We employ a single-layer kagome lattice (see Fig.~\ref{fig:system}) model with ferromagnetic order.
 The tight-binding Hamiltonian is,
\begin{align} \label{Kagome_hamiltonian}
 H         = {}&  t\sum_{ \braket{i,j} } c^\dagger_i \sigma_0 c_j
              + \mathrm{i} \lambda\_{SO}\sum_{ \braket{\braket{i,j}} } c^\dagger_{i} \nu_{i,j} \sigma_z c_{j}  \nonumber\\
            {}& + J \sum_i c^\dagger_i \bm{M}_i \cdot \bm{\sigma}c_i + \sum_i c^\dagger_i V_i \sigma_0 c_i\,.
\end{align}
 The first term is the nearest-neighbor hopping.
 The second term is the spin-orbit coupling conserving $z$-component of electron spin \cite{Kane05z2,Guo09topological},
where $\nu_{ij} = \frac{2}{\sqrt{3}}(\bm{d}_{im} \times \bm{d}_{mj}) \cdot \bm{e}_z = \pm 1$ and 
$\bm{d}_{im}$ is the unit vector connecting the sites $i$ and $m$,
with $m$ the site in between next-nearest sites $i$ and $j$.
 The third term is the spin-exchange coupling with local magnetization $\bm{M}_i$.
The magnetization comes from localized spins for high-spin systems (e.g.~Fe$_3$Sn$_2$ \cite{Caer78mossbauer, Ye18massive, Yin18giant})  
and from the mean field of itinerant spins for low-spin systems (e.g.~Co$_3$Sn$_2$S$_2$ \cite{Liu18giant, Ozawa19two}).
 The fourth term is the on-site random potential (non-magnetic disorder),
which is uniformly distributed in $[-{W\over 2},{W\over 2}]$.
Pauli matrices $\bm \sigma$ represent the spin degree of freedom.
 We take the hopping parameter $t$ as the energy unit and
the distance $a$ between the nearest-neighbor sites as the length unit.
 The strength of the spin-orbit coupling is set to $\lambda\_{SO} = 0.5t$ in the following.

 We consider straight-edged nanoribbons of the kagome layer (Fig.~\ref{fig:system}), with length $L_x=(5N-1)a$, and width $L_y=\frac{\sqrt{3}}{2}(N-1)a$.
 We calculate the two-terminal conductance $G$ (in units of $e^2/h$) between the terminals attached on the ends of the ribbon, $x=0$ and $x=L_x$, by using the recursive Green's function method \cite{Ando91quantum}.
 Since we found that the system size dependence is not important for the qualitative behavior of DWMR,
we show only the data for $N=31$ here.

\begin{figure}[tbp]
 \centering
  \includegraphics[width=\linewidth]{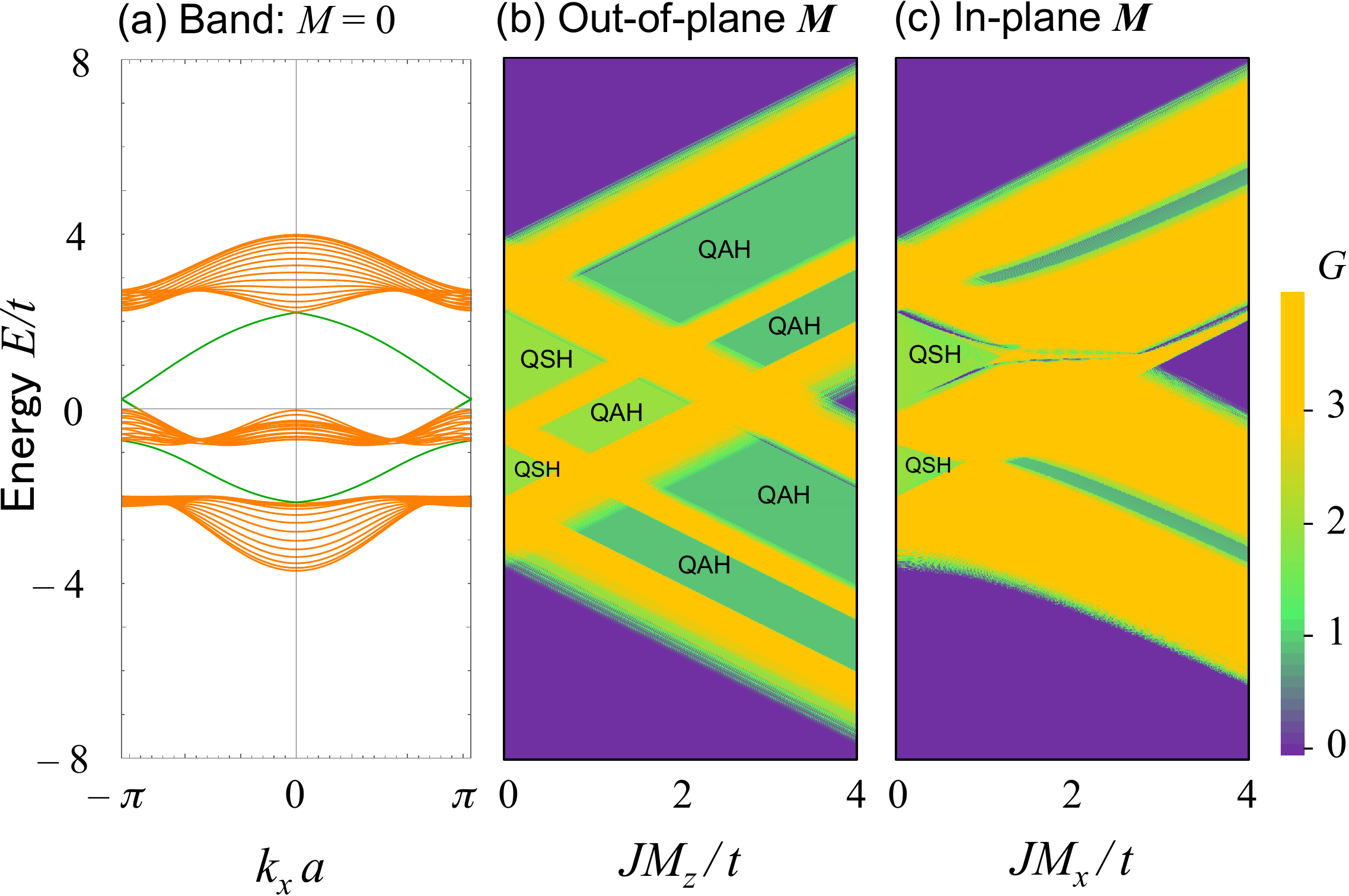}
 \vspace{-5mm}
\caption{%
  (a) Band structure of a non-magnetic kagome ribbon.
  Green bands correspond to the helical edge states.
  Conductance maps for clean kagome nanoribbons with (b) out-of-plane and (c) in-plane magnetizations.
  The vertical axis is the Fermi energy $E$, and the horizontal axis is the strength of the spin-exchange coupling $JM$.
  The light-green regions correspond to the quantum spin Hall phase (near $JM = 0$) or the QAH phase with Chern number $2$.
  The dark-green regions in (b) are the QAH phase with Chern number $\pm 1$.
}
\label{fig:BandGmap}
\end{figure}

\begin{figure}[tbp]
 \centering
  \includegraphics[width=\linewidth]{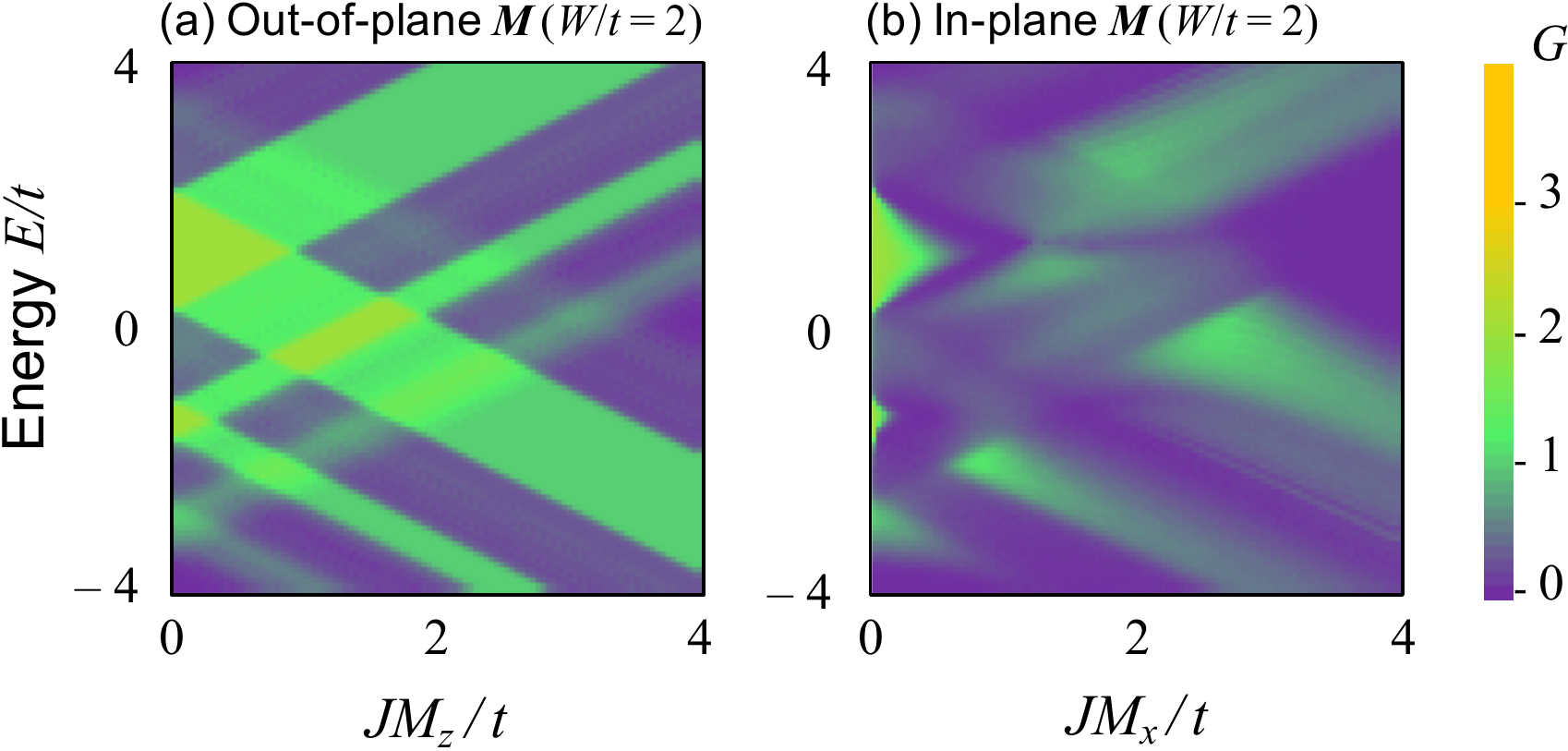}
 \vspace{-5mm}
\caption{%
  Conductance maps for disordered (with disorder strength $W/t=2$) kagome nanoribbons with (a) out-of-plane and (b) in-plane magnetizations.
  The metallic (yellow) regions in Fig.~\ref{fig:BandGmap} become insulating (purple) in the presence of disorder.
  We took the disorder realization average over $10^3$ samples.
} \vspace{2mm}
\label{fig:mapDisordered}
\end{figure}

\textit{Two-terminal conductance.}
 We first study the transport with uniform magnetization.
 Without magnetization, the kagome lattice system shows a quantum spin Hall state \cite{Guo09topological}, similarly to the honeycomb lattice with spin-orbit coupling.
 Under strong spin-exchange coupling, qualitatively different behavior arises depending on the magnetization direction [Figs.~\ref{fig:BandGmap}(b) and \ref{fig:BandGmap}(c)].
 For out-of-plane ($z$) magnetizations, the system shows the QAH states,
where quantized conductance arises due to the chiral edge states.
 (Note that the quantum spin Hall state survives for a small $M_z$ because the $\sigma_z$ term does not break the symmetry of the Hamiltonian.)
 In contrast, for in-plane ($x$ or $y$) magnetizations,
 the system tends to be ``diffusive'':
metallic in the clean limit and insulating
in the presence of disorder, due to the Anderson localization (see Fig.~\ref{fig:mapDisordered}).
 We can switch the type of transport, QAH and diffusive, by tilting the magnetization \cite{Kandala15giant,Kou15metal},
and the difference is significantly contrasted in the presence of disorder.

\begin{figure}[tbp]
 \centering
  \includegraphics[width=0.98\linewidth]{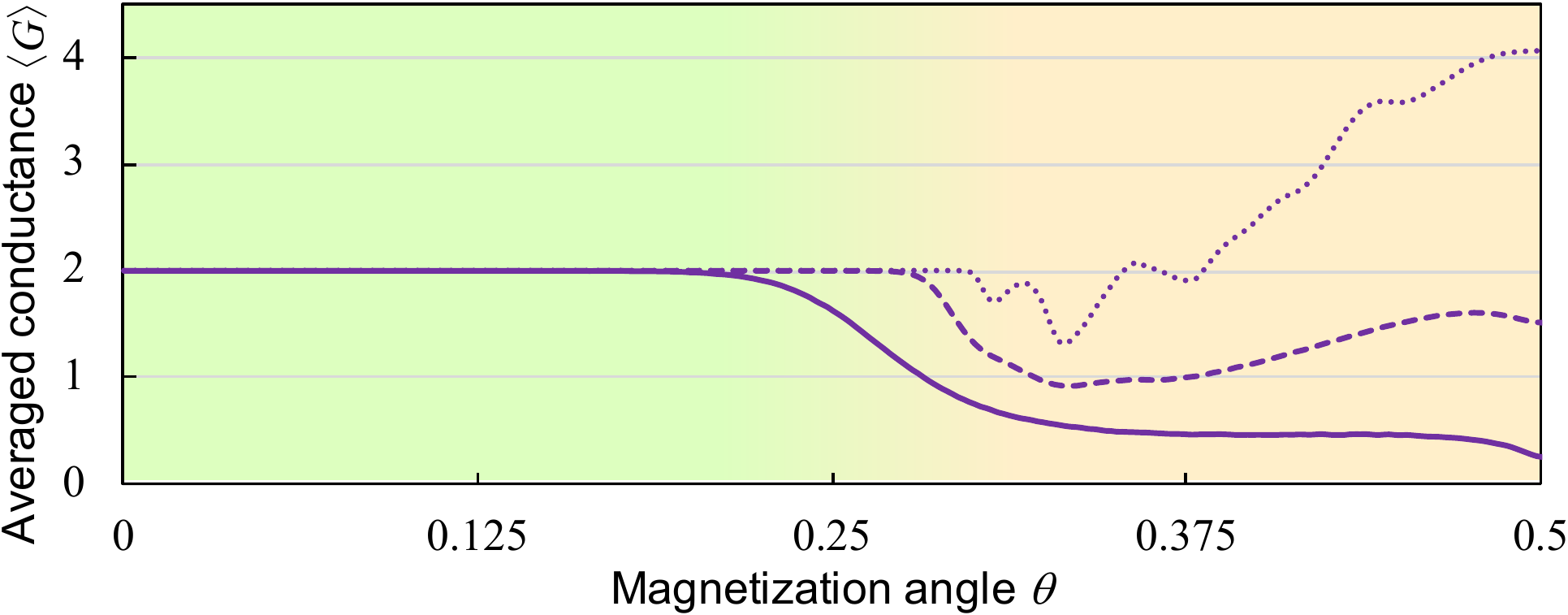}
 \vspace{-2mm}
\caption{%
  Conductance as a function of the angle $\theta$ (in units of $\pi$) of magnetization $\bm{M} = M(\sin\theta,0,\cos\theta)$ at $W/t=0$ (dotted), $1$ (dashed), and $2$ (solid).
  The parameters are set to $JM/t=1$ and $E/t=-0.4$, which is around the center of the QAH region with Chern number $2$ in Fig.~\ref{fig:BandGmap}(b).
  Averages are taken over up to $10^4$ samples, so that the error bars are smaller than $0.005 e^2/h$.
}
\label{fig:thetaDep}
\end{figure}

 Next we show the magnetization angle $\theta$ dependence of conductance (Fig.~\ref{fig:thetaDep}).
 For a small $\theta$, where the magnetization is almost in the out-of-plane direction, 
the conductance is quantized even in the presence of disorder.
 This implies that the system is in the QAH state and shows the edge transport.
 For a large $\theta$, where the in-plane component of the magnetization $M_x$ becomes large, 
the system shows conducting or insulating behavior depending on the disorder strength and system size.
 This is the feature of the diffusive transport.
 The QAH-diffusive crossover occurs when the bulk band touches the Fermi energy.
 This crossover is understood as the competition of the spin-orbit term $\lambda\_{SO}$ 
and in-plane component of the magnetization $J M_x$,
which opens and narrows the bulk band gap, respectively.
 This magnetization-angle induced change of transport property leads to an unconventional type of magnetoresistance effect by spatial modulation of the magnetic texture: domain walls.

\textit{Domain-wall magnetoresistance.}
 Then we study the transport in systems with domain walls.
 We consider three types of domain walls: N{\'e}el, head-to-head, and in-plane.
 Note that a Bloch-type wall gives the same results as a N{\'e}el type
because the in-plane magnetizations, $\sigma_x$ and $\sigma_y$, are equivalent in our model.
 The domain walls of thickness $\xi$ can be implemented by position $x$ dependent magnetizations,
\begin{align}
 \bm{M}(x)= M \( \mathrm{sech} \frac{x\!-\!L_x/2}{\xi}, 0, - \tanh \frac{x\!-\!L_x/2}{\xi} \),
 \label{eqn:Neel}
\end{align}
for a N{\'e}el wall,
\begin{align}
 \bm{M}(x)= M \( - \tanh \frac{x\!-\!L_x/2}{\xi}, 0, \mathrm{sech} \frac{x\!-\!L_x/2}{\xi} \),
 \label{eqn:h2h}
\end{align}
for a head-to-head wall, and 
\begin{align}
 \bm{M}(x)= M \( \mathrm{sech} \frac{x\!-\!L_x/2}{\xi}, - \tanh \frac{x\!-\!L_x/2}{\xi}, 0 \),
 \label{eqn:inPlaneWall}
\end{align}
for an in-plane wall.

\begin{figure}[tbp]
  \includegraphics[width=1\linewidth]{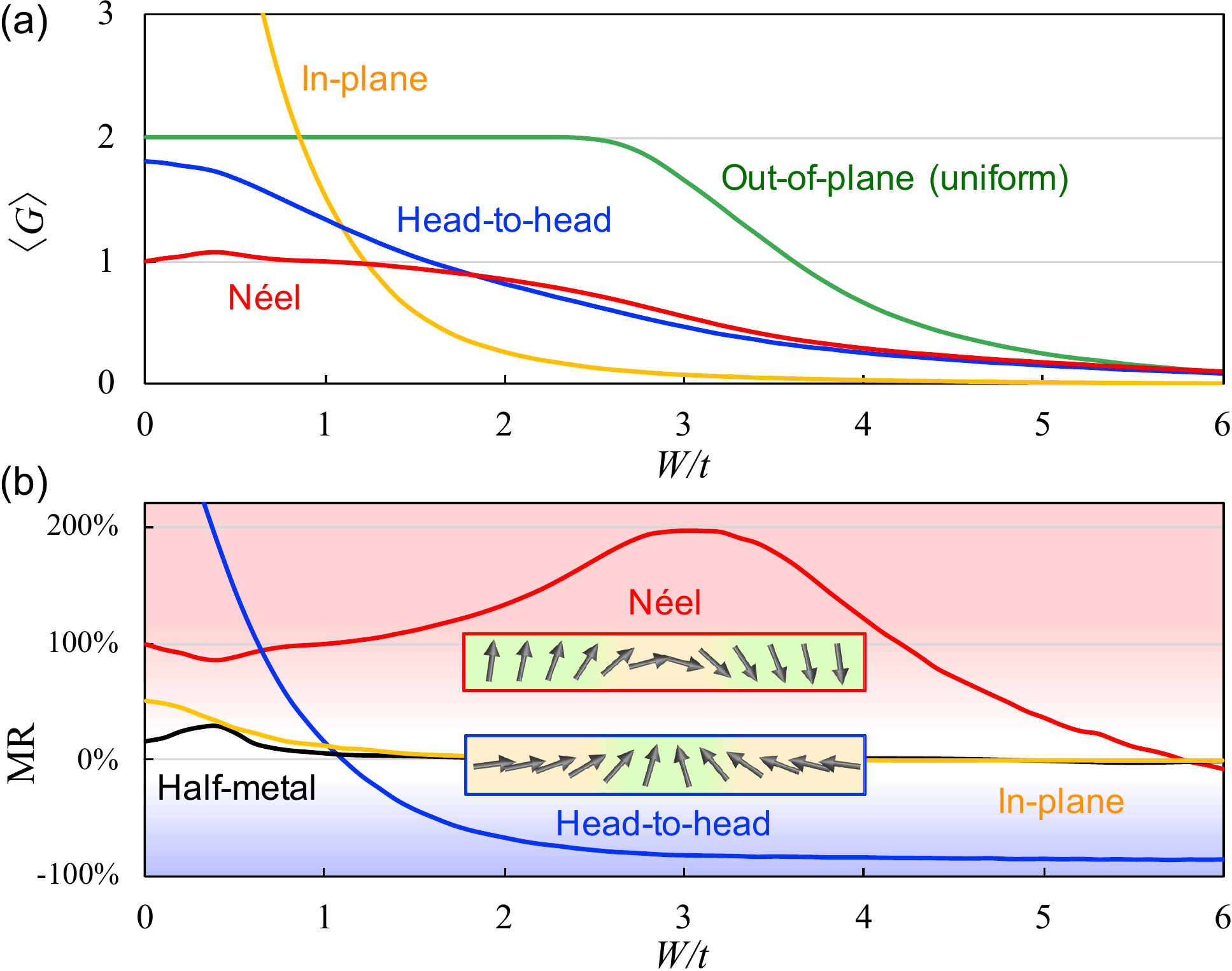}
 \vspace{-3mm}
\caption{%
  (a) Averaged conductance $\braket{G}$ for (green) out-of-plane magnetization, 
  (yellow) in-plane magnetization, 
  (red) N{\'e}el, and 
  (blue) head-to-head walls with the domain-wall thickness $\xi/a=30$.
  The parameters are set to $JM/t=1$ and $E/t=-0.4$.
  (b) Magnetoresistance ratio MR for (red) N{\'e}el, (blue) head-to-head, and (yellow) in-plane walls with $\xi/a=0.5$.
   The black line shows the MR in a conventional half-metal state (at $JM/t=1$ and $E/t=4$) for N{\'e}el wall with $\xi/a=0.5$.
}
\label{fig:MR}
\end{figure}

 Figure~\ref{fig:MR}(a) shows the averaged conductance with uniform magnetizations or domain walls under disorder.
 The conductance for the out-of-plane uniform magnetization shows that the QAH state (quantized plateau) breaks down around $W/t=2.8$
in the case of $JM/t=1$ and $E/t=-0.4$.
 Under the in-plane magnetization (either uniform or domain wall), 
where the disordered system shows diffusive transport,
the conductance quickly decays as the disorder strength increases.
 On the other hand, the conductance with a N{\'e}el or head-to-head domain wall shows an intermediate behavior against disorder.

 The DWMR effect is characterized by the magnetoresistance ratio (MR),
which is defined by the ratio of the conductance in uniform magnetization $G\_{uni}$ to that in a domain wall $G\_{DW}$ as
\begin{align}
 \textrm{MR} &= \frac{\braket{G\_{uni}}}{\braket{G\_{DW}}}-1,
 \label{eqn:MR}
\end{align}
where $\braket{\cdots}$ represents the average over disorder realizations.
 Figure~\ref{fig:MR}(b) shows the domain-wall MR as functions of disorder strength $W$.
 The MR for the N{\'e}el wall, where $G\_{uni}=G\_{out-of-plane}$ and $G\_{DW}=G\_{N{\'e}el}$, is significantly large (about $100\%$) at a weak disorder.
 It shows a maximum (about $200\%$) at an intermediate disorder strength.
 The position of the maxima ($W/t\simeq 2.8$) corresponds to the point where disorder-driven QAH-diffusive crossover occurs.
 After the transition, the system shows diffusive transport in both the in-plane and out-of-plane magnetizations, and the MR vanishes.
 This large MR in the N{\'e}el wall arises because 
the diffusive region located around the domain wall works as a resistor in the disorder tolerant QAH states.
 The MR for the head-to-head wall, where $G\_{uni}=G\_{in-plane}$ and $G\_{DW}=G\_{head-to-head}$, 
becomes negative at a strong disorder, while it is positive and large for a weak disorder.
 The sign change of the MR occurs when the conductance in the diffusive state becomes comparable with that in the QAH state.
 Although the MR keeps negative value for a strong disorder, 
the difference $G\_{head-to-head} - G\_{in-plane}$ is maximized around the phase transition point $W/t\simeq 2.8$.
 For an in-plane wall, the conductance shows almost the same behavior as the case of uniform magnetization.
 Thus the MR is vanishing in disordered systems, as in conventional half-metals.
 In materials with an in-plane easy axis, the in-plane type wall is more favored than the head-to-head type that has the out-of-plane magnetization near the center of the wall,
and the huge negative magnetoresistance may be difficult to observe.
 Therefore, it should be easier to obtain the huge positive magnetoresistance of N{\'e}el/Bloch walls in the kagome materials with out-of-plane easy axes, such as thin films of Co$_3$Sn$_2$S$_2$.

\begin{figure}[tbp]
  \includegraphics[width=1\linewidth]{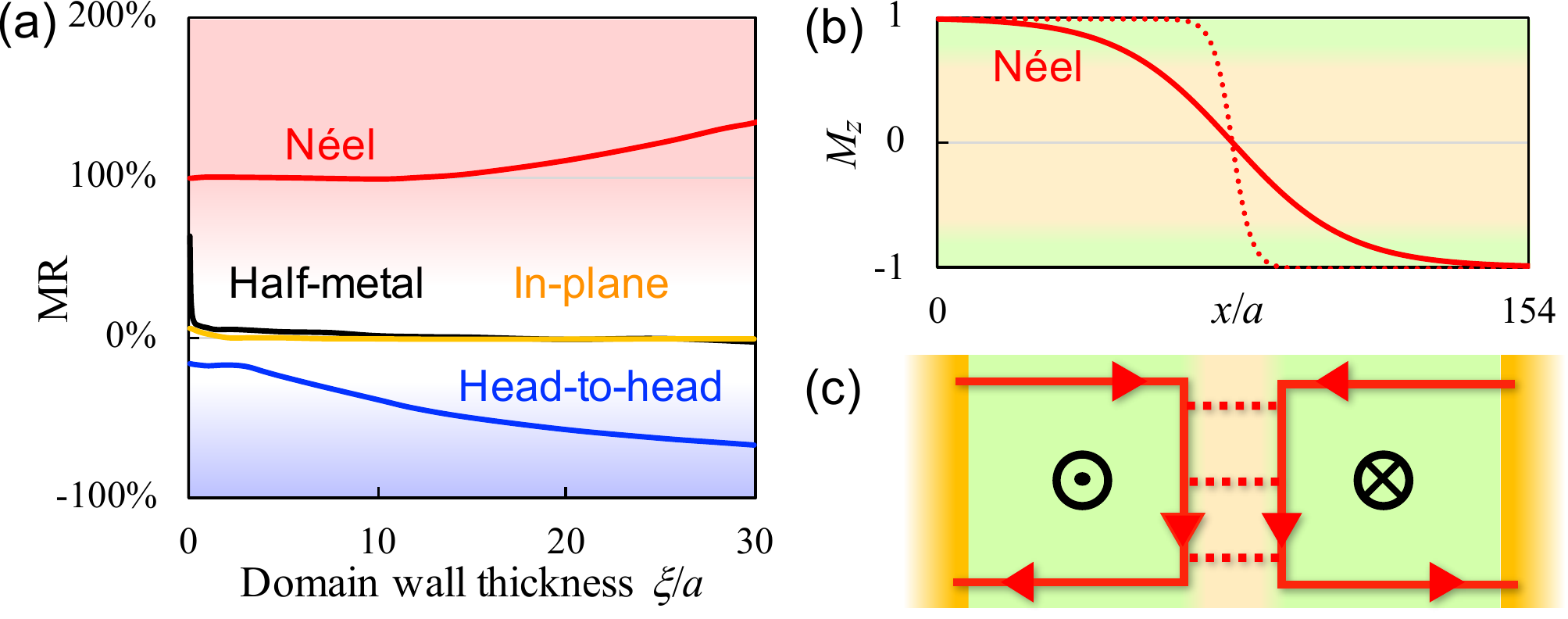}
 \vspace{-2mm}
\caption{%
  (a) Magnetoresistance ratio MR as a function of domain-wall thickness for 
  (red) N{\'e}el, (blue) head-to-head, and (yellow) in-plane walls.
  The black line shows the conventional DWMR effect in a half-metal, which vanishes in a thick wall.
  The disorder strength $W/t=2$.
  (b) Spatial configuration of an out-of-plane component of magnetization $M_z$ for a N{\'e}el wall with (solid) $\xi/a=30$ and (dotted) $\xi/a=5$.
  (c) Schematic figure of the chiral edge state coupled at the domain wall.
}
\label{fig:thickness}
\end{figure}

 Lastly, we study the effect of domain-wall thickness on the MR (see Fig.~\ref{fig:thickness}).
 The magnetoresistance effect in kagome layers with N{\'e}el or head-to-head walls is enhanced as the thickness increases.
 This is in good contrast with a conventional DWMR in half-metals, which quickly vanishes for increasing domain-wall thickness, irrespective of the wall type;
the spin of the conduction electron can change gradually and the current can go through the domain walls.
 At the center of the domain walls, the magnetization is $\pi/2$ rotated with respect to the background.
 Therefore the transport properties in both ends (deeply in the domains) and near the center of domain wall are significantly different (diffusive/QAH transport) as shown in Fig.~\ref{fig:thetaDep};
in the presence of disorder, the transport is suppressed in the in-plane magnetized (diffusive) region and enhanced in the out-of-plane magnetized (QAH) region.
 Therefore the MRs are magnified as the wall (where the transport property changes) thickness increases.
 We also note that the MR for a thin N{\'e}el wall is $100\%$, 
because the edge states mix at the domain wall and a half of them go through the wall [see Fig.~\ref{fig:thickness}(c)].
 The transport in QAH systems with domain-wall resistance is also studied by experiments \cite{Yasuda17quantized}.
 If there are multiple walls (such as a racetrack memory), the resistance additively increases and the MR becomes huge.

\textit{Conclusion.}
 We have studied the transport in disordered ferromagnetic kagome layers.
 We found a huge and stable DWMR effect originating from the chiral edge states of hte QAH system and the magnetization induced topological phase transition.
 Considering N{\'e}el or Bloch domain walls, 
we have shown that a huge (around $100\%$) MR is achieved, irrespective of the domain-wall thickness and weak disorder strength.
 We have also shown that a negative DWMR can be realized in the head-to-head wall.
 These features are contrasted with the conventional DWMR in half-metals, which is always positive, fragile against disorder, and vanishingly small in thick walls.
 This robust magnetoresistance in QAH kagome layers will make an opportunity to realize the disorder-tolerant and low-power consumption devices.
 For instance, the current and Joule heating in the racetrack memory can be suppressed by using 
ferromagnetic kagome layers, such as thin films of Co$_3$Sn$_2$S$_2$.
 We also note that these topologically protected DWMR effects will pave the way to realize single-material spintronics devices.
 The absence/presence of a domain wall can be used as ``0''/``1''.
 Furthermore, by simply changing the number of domain walls, 
we may realize an analog-non-volatile memory, which is now longed for in neuromorphic computing and deep learning.
 We expect the experimental realization of domain walls in kagome thin-films is not difficult, 
since the Curie temperature is high enough, $T\_C=657\,$K \cite{Caer78mossbauer} for Fe$_3$Sn$_2$ 
and $T\_C=175\,$K \cite{Liu18giant} for Co$_3$Sn$_2$S$_2$.
 The domains can be engineered by using a junction of different coercivity or just \textit{writing} the domains \cite{Yasuda17quantized}.


\textit{Acknowledgments.}
 We thank Tomi Ohtsuki, Yuya Ominato, and Mai Kameda for valuable discussions.
 This work was supported by the Japan Society for the Promotion of Science 
KAKENHI (Grant Nos. 
JP15H05854, 
JP16J01981, 
JP17K05485, and  
JP19K14607), 
and 
by CREST, Japan Science and Technology Agency (Grant No.~JPMJCR18T2).

\bibliography{Takagome}

\begin{thebibliography}{35}%
\makeatletter
\providecommand \@ifxundefined [1]{%
 \@ifx{#1\undefined}
}%
\providecommand \@ifnum [1]{%
 \ifnum #1\expandafter \@firstoftwo
 \else \expandafter \@secondoftwo
 \fi
}%
\providecommand \@ifx [1]{%
 \ifx #1\expandafter \@firstoftwo
 \else \expandafter \@secondoftwo
 \fi
}%
\providecommand \natexlab [1]{#1}%
\providecommand \enquote  [1]{``#1''}%
\providecommand \bibnamefont  [1]{#1}%
\providecommand \bibfnamefont [1]{#1}%
\providecommand \citenamefont [1]{#1}%
\providecommand \href@noop [0]{\@secondoftwo}%
\providecommand \href [0]{\begingroup \@sanitize@url \@href}%
\providecommand \@href[1]{\@@startlink{#1}\@@href}%
\providecommand \@@href[1]{\endgroup#1\@@endlink}%
\providecommand \@sanitize@url [0]{\catcode `\\12\catcode `\$12\catcode
  `\&12\catcode `\#12\catcode `\^12\catcode `\_12\catcode `\%12\relax}%
\providecommand \@@startlink[1]{}%
\providecommand \@@endlink[0]{}%
\providecommand \url  [0]{\begingroup\@sanitize@url \@url }%
\providecommand \@url [1]{\endgroup\@href {#1}{\urlprefix }}%
\providecommand \urlprefix  [0]{URL }%
\providecommand \Eprint [0]{\href }%
\providecommand \doibase [0]{http://dx.doi.org/}%
\providecommand \selectlanguage [0]{\@gobble}%
\providecommand \bibinfo  [0]{\@secondoftwo}%
\providecommand \bibfield  [0]{\@secondoftwo}%
\providecommand \translation [1]{[#1]}%
\providecommand \BibitemOpen [0]{}%
\providecommand \bibitemStop [0]{}%
\providecommand \bibitemNoStop [0]{.\EOS\space}%
\providecommand \EOS [0]{\spacefactor3000\relax}%
\providecommand \BibitemShut  [1]{\csname bibitem#1\endcsname}%
\let\auto@bib@innerbib\@empty
\bibitem [{\citenamefont {Parkin}\ \emph {et~al.}(2008)\citenamefont {Parkin},
  \citenamefont {Hayashi},\ and\ \citenamefont {Thomas}}]{Parkin08magnetic}%
  \BibitemOpen
  \bibfield  {author} {\bibinfo {author} {\bibfnamefont {S.~S.~P.}\
  \bibnamefont {Parkin}}, \bibinfo {author} {\bibfnamefont {M.}~\bibnamefont
  {Hayashi}}, \ and\ \bibinfo {author} {\bibfnamefont {L.}~\bibnamefont
  {Thomas}},\ }\href@noop {} {\bibfield  {journal} {\bibinfo  {journal}
  {Science}\ }\textbf {\bibinfo {volume} {320}},\ \bibinfo {pages} {190}
  (\bibinfo {year} {2008})}\BibitemShut {NoStop}%
\bibitem [{\citenamefont {Parkin}(1995)}]{Parkin95giant}%
  \BibitemOpen
  \bibfield  {author} {\bibinfo {author} {\bibfnamefont {S.~S.~P.}\
  \bibnamefont {Parkin}},\ }\href@noop {} {\bibfield  {journal} {\bibinfo
  {journal} {Annu. Rev. Mater. Sci.}\ }\textbf {\bibinfo {volume} {25}},\
  \bibinfo {pages} {357} (\bibinfo {year} {1995})}\BibitemShut {NoStop}%
\bibitem [{\citenamefont {Haldane}(1988)}]{Haldane88model}%
  \BibitemOpen
  \bibfield  {author} {\bibinfo {author} {\bibfnamefont {F.~D.~M.}\
  \bibnamefont {Haldane}},\ }\href@noop {} {\bibfield  {journal} {\bibinfo
  {journal} {Phys. Rev. Lett.}\ }\textbf {\bibinfo {volume} {61}},\ \bibinfo
  {pages} {2015} (\bibinfo {year} {1988})}\BibitemShut {NoStop}%
\bibitem [{\citenamefont {Chang}\ \emph {et~al.}(2013)\citenamefont {Chang},
  \citenamefont {Zhang}, \citenamefont {Feng}, \citenamefont {Shen},
  \citenamefont {Zhang}, \citenamefont {Guo}, \citenamefont {Li}, \citenamefont
  {Ou}, \citenamefont {Wei}, \citenamefont {Wang}, \citenamefont {Ji},
  \citenamefont {Feng}, \citenamefont {Ji}, \citenamefont {Chen}, \citenamefont
  {Jia}, \citenamefont {Dai}, \citenamefont {Fang}, \citenamefont {Zhang},
  \citenamefont {He}, \citenamefont {Wang}, \citenamefont {Lu}, \citenamefont
  {Ma},\ and\ \citenamefont {Xue}}]{Chang13experimental}%
  \BibitemOpen
  \bibfield  {author} {\bibinfo {author} {\bibfnamefont {C.-Z.}\ \bibnamefont
  {Chang}}, \bibinfo {author} {\bibfnamefont {J.}~\bibnamefont {Zhang}},
  \bibinfo {author} {\bibfnamefont {X.}~\bibnamefont {Feng}}, \bibinfo {author}
  {\bibfnamefont {J.}~\bibnamefont {Shen}}, \bibinfo {author} {\bibfnamefont
  {Z.}~\bibnamefont {Zhang}}, \bibinfo {author} {\bibfnamefont
  {M.}~\bibnamefont {Guo}}, \bibinfo {author} {\bibfnamefont {K.}~\bibnamefont
  {Li}}, \bibinfo {author} {\bibfnamefont {Y.}~\bibnamefont {Ou}}, \bibinfo
  {author} {\bibfnamefont {P.}~\bibnamefont {Wei}}, \bibinfo {author}
  {\bibfnamefont {L.-L.}\ \bibnamefont {Wang}}, \bibinfo {author}
  {\bibfnamefont {Z.-Q.}\ \bibnamefont {Ji}}, \bibinfo {author} {\bibfnamefont
  {Y.}~\bibnamefont {Feng}}, \bibinfo {author} {\bibfnamefont {S.}~\bibnamefont
  {Ji}}, \bibinfo {author} {\bibfnamefont {X.}~\bibnamefont {Chen}}, \bibinfo
  {author} {\bibfnamefont {J.}~\bibnamefont {Jia}}, \bibinfo {author}
  {\bibfnamefont {X.}~\bibnamefont {Dai}}, \bibinfo {author} {\bibfnamefont
  {Z.}~\bibnamefont {Fang}}, \bibinfo {author} {\bibfnamefont {S.-C.}\
  \bibnamefont {Zhang}}, \bibinfo {author} {\bibfnamefont {K.}~\bibnamefont
  {He}}, \bibinfo {author} {\bibfnamefont {Y.}~\bibnamefont {Wang}}, \bibinfo
  {author} {\bibfnamefont {L.}~\bibnamefont {Lu}}, \bibinfo {author}
  {\bibfnamefont {X.-C.}\ \bibnamefont {Ma}}, \ and\ \bibinfo {author}
  {\bibfnamefont {Q.-K.}\ \bibnamefont {Xue}},\ }\href@noop {} {\bibfield
  {journal} {\bibinfo  {journal} {Science}\ }\textbf {\bibinfo {volume}
  {340}},\ \bibinfo {pages} {167} (\bibinfo {year} {2013})}\BibitemShut
  {NoStop}%
\bibitem [{\citenamefont {Wan}\ \emph {et~al.}(2011)\citenamefont {Wan},
  \citenamefont {Turner}, \citenamefont {Vishwanath},\ and\ \citenamefont
  {Savrasov}}]{Wan11}%
  \BibitemOpen
  \bibfield  {author} {\bibinfo {author} {\bibfnamefont {X.}~\bibnamefont
  {Wan}}, \bibinfo {author} {\bibfnamefont {A.~M.}\ \bibnamefont {Turner}},
  \bibinfo {author} {\bibfnamefont {A.}~\bibnamefont {Vishwanath}}, \ and\
  \bibinfo {author} {\bibfnamefont {S.~Y.}\ \bibnamefont {Savrasov}},\
  }\href@noop {} {\bibfield  {journal} {\bibinfo  {journal} {Phys. Rev. B}\
  }\textbf {\bibinfo {volume} {83}},\ \bibinfo {pages} {205101} (\bibinfo
  {year} {2011})}\BibitemShut {NoStop}%
\bibitem [{\citenamefont {Burkov}\ and\ \citenamefont
  {Balents}(2011)}]{Burkov11}%
  \BibitemOpen
  \bibfield  {author} {\bibinfo {author} {\bibfnamefont {A.~A.}\ \bibnamefont
  {Burkov}}\ and\ \bibinfo {author} {\bibfnamefont {L.}~\bibnamefont
  {Balents}},\ }\href@noop {} {\bibfield  {journal} {\bibinfo  {journal} {Phys.
  Rev. Lett.}\ }\textbf {\bibinfo {volume} {107}},\ \bibinfo {pages} {127205}
  (\bibinfo {year} {2011})}\BibitemShut {NoStop}%
\bibitem [{\citenamefont {Upadhyaya}\ and\ \citenamefont
  {Tserkovnyak}(2016)}]{Upadhyaya16domain}%
  \BibitemOpen
  \bibfield  {author} {\bibinfo {author} {\bibfnamefont {P.}~\bibnamefont
  {Upadhyaya}}\ and\ \bibinfo {author} {\bibfnamefont {Y.}~\bibnamefont
  {Tserkovnyak}},\ }\href@noop {} {\bibfield  {journal} {\bibinfo  {journal}
  {Phys. Rev. B}\ }\textbf {\bibinfo {volume} {94}},\ \bibinfo {pages}
  {020411(R)} (\bibinfo {year} {2016})}\BibitemShut {NoStop}%
\bibitem [{\citenamefont {Araki}\ \emph {et~al.}(2016)\citenamefont {Araki},
  \citenamefont {Yoshida},\ and\ \citenamefont {Nomura}}]{Araki16universal}%
  \BibitemOpen
  \bibfield  {author} {\bibinfo {author} {\bibfnamefont {Y.}~\bibnamefont
  {Araki}}, \bibinfo {author} {\bibfnamefont {A.}~\bibnamefont {Yoshida}}, \
  and\ \bibinfo {author} {\bibfnamefont {K.}~\bibnamefont {Nomura}},\
  }\href@noop {} {\bibfield  {journal} {\bibinfo  {journal} {Phys. Rev. B}\
  }\textbf {\bibinfo {volume} {94}},\ \bibinfo {pages} {115312} (\bibinfo
  {year} {2016})}\BibitemShut {NoStop}%
\bibitem [{\citenamefont {Kurebayashi}\ and\ \citenamefont
  {Nomura}(2019)}]{Kurebayashi19theory}%
  \BibitemOpen
  \bibfield  {author} {\bibinfo {author} {\bibfnamefont {D.}~\bibnamefont
  {Kurebayashi}}\ and\ \bibinfo {author} {\bibfnamefont {K.}~\bibnamefont
  {Nomura}},\ }\href@noop {} {\bibfield  {journal} {\bibinfo  {journal} {Sci.
  Rep.}\ }\textbf {\bibinfo {volume} {9}},\ \bibinfo {pages} {5365} (\bibinfo
  {year} {2019})}\BibitemShut {NoStop}%
\bibitem [{\citenamefont {Kim}\ \emph {et~al.}(2019)\citenamefont {Kim},
  \citenamefont {Kurebayashi},\ and\ \citenamefont
  {Nomura}}]{Kim19electrically}%
  \BibitemOpen
  \bibfield  {author} {\bibinfo {author} {\bibfnamefont {S.}~\bibnamefont
  {Kim}}, \bibinfo {author} {\bibfnamefont {D.}~\bibnamefont {Kurebayashi}}, \
  and\ \bibinfo {author} {\bibfnamefont {K.}~\bibnamefont {Nomura}},\
  }\href@noop {} {\bibfield  {journal} {\bibinfo  {journal} {J. Phys. Soc.
  Jpn.}\ }\textbf {\bibinfo {volume} {88}},\ \bibinfo {pages} {083704}
  (\bibinfo {year} {2019})}\BibitemShut {NoStop}%
\bibitem [{\citenamefont {Kent}\ \emph {et~al.}(2001)\citenamefont {Kent},
  \citenamefont {Yu}, \citenamefont {R\"udiger},\ and\ \citenamefont
  {Parkin}}]{Kent01domain}%
  \BibitemOpen
  \bibfield  {author} {\bibinfo {author} {\bibfnamefont {A.~D.}\ \bibnamefont
  {Kent}}, \bibinfo {author} {\bibfnamefont {J.}~\bibnamefont {Yu}}, \bibinfo
  {author} {\bibfnamefont {U.}~\bibnamefont {R\"udiger}}, \ and\ \bibinfo
  {author} {\bibfnamefont {S.~S.~P.}\ \bibnamefont {Parkin}},\ }\href@noop {}
  {\bibfield  {journal} {\bibinfo  {journal} {J. Phys.: Condens. Matter}\
  }\textbf {\bibinfo {volume} {13}},\ \bibinfo {pages} {R461} (\bibinfo {year}
  {2001})}\BibitemShut {NoStop}%
\bibitem [{\citenamefont {Maekawa}\ \emph {et~al.}(2012)\citenamefont
  {Maekawa}, \citenamefont {Valenzuela}, \citenamefont {Saitoh},\ and\
  \citenamefont {Kimura}}]{MaekawaBook}%
  \BibitemOpen
  \bibfield  {author} {\bibinfo {author} {\bibfnamefont {S.}~\bibnamefont
  {Maekawa}}, \bibinfo {author} {\bibfnamefont {S.~O.}\ \bibnamefont
  {Valenzuela}}, \bibinfo {author} {\bibfnamefont {E.}~\bibnamefont {Saitoh}},
  \ and\ \bibinfo {author} {\bibfnamefont {T.}~\bibnamefont {Kimura}},\
  }\href@noop {} {\emph {\bibinfo {title} {{Spin Current (Semiconductor Science
  and Technology)}}}}\ (\bibinfo  {publisher} {Oxford University Press, Oxford,
  U.K.},\ \bibinfo {year} {2012})\BibitemShut {NoStop}%
\bibitem [{\citenamefont {Yavorsky}\ \emph {et~al.}(2002)\citenamefont
  {Yavorsky}, \citenamefont {Mertig}, \citenamefont {Perlov}, \citenamefont
  {Yaresko},\ and\ \citenamefont {Antonov}}]{Yavorsky02giant}%
  \BibitemOpen
  \bibfield  {author} {\bibinfo {author} {\bibfnamefont {B.~Y.}\ \bibnamefont
  {Yavorsky}}, \bibinfo {author} {\bibfnamefont {I.}~\bibnamefont {Mertig}},
  \bibinfo {author} {\bibfnamefont {A.~Y.}\ \bibnamefont {Perlov}}, \bibinfo
  {author} {\bibfnamefont {A.~N.}\ \bibnamefont {Yaresko}}, \ and\ \bibinfo
  {author} {\bibfnamefont {V.~N.}\ \bibnamefont {Antonov}},\ }\href@noop {}
  {\bibfield  {journal} {\bibinfo  {journal} {Phys. Rev. B}\ }\textbf {\bibinfo
  {volume} {66}},\ \bibinfo {pages} {174422} (\bibinfo {year}
  {2002})}\BibitemShut {NoStop}%
\bibitem [{\citenamefont {Hirschberger}\ \emph {et~al.}(2016)\citenamefont
  {Hirschberger}, \citenamefont {Kushwaha}, \citenamefont {Wang}, \citenamefont
  {Gibson}, \citenamefont {Liang}, \citenamefont {Belvin}, \citenamefont
  {Bernevig}, \citenamefont {Cava},\ and\ \citenamefont
  {Ong}}]{Hirschberger16}%
  \BibitemOpen
  \bibfield  {author} {\bibinfo {author} {\bibfnamefont {M.}~\bibnamefont
  {Hirschberger}}, \bibinfo {author} {\bibfnamefont {S.}~\bibnamefont
  {Kushwaha}}, \bibinfo {author} {\bibfnamefont {Z.}~\bibnamefont {Wang}},
  \bibinfo {author} {\bibfnamefont {Q.}~\bibnamefont {Gibson}}, \bibinfo
  {author} {\bibfnamefont {S.}~\bibnamefont {Liang}}, \bibinfo {author}
  {\bibfnamefont {C.~A.}\ \bibnamefont {Belvin}}, \bibinfo {author}
  {\bibfnamefont {B.~A.}\ \bibnamefont {Bernevig}}, \bibinfo {author}
  {\bibfnamefont {R.~J.}\ \bibnamefont {Cava}}, \ and\ \bibinfo {author}
  {\bibfnamefont {N.~P.}\ \bibnamefont {Ong}},\ }\href@noop {} {\bibfield
  {journal} {\bibinfo  {journal} {Nat. Mater.}\ }\textbf {\bibinfo {volume}
  {15}},\ \bibinfo {pages} {1161} (\bibinfo {year} {2016})}\BibitemShut
  {NoStop}%
\bibitem [{\citenamefont {Wang}\ \emph {et~al.}(2016)\citenamefont {Wang},
  \citenamefont {Vergniory}, \citenamefont {Kushwaha}, \citenamefont
  {Hirschberger}, \citenamefont {Chulkov}, \citenamefont {Ernst}, \citenamefont
  {Ong}, \citenamefont {Cava},\ and\ \citenamefont {Bernevig}}]{Wang16}%
  \BibitemOpen
  \bibfield  {author} {\bibinfo {author} {\bibfnamefont {Z.}~\bibnamefont
  {Wang}}, \bibinfo {author} {\bibfnamefont {M.~G.}\ \bibnamefont {Vergniory}},
  \bibinfo {author} {\bibfnamefont {S.}~\bibnamefont {Kushwaha}}, \bibinfo
  {author} {\bibfnamefont {M.}~\bibnamefont {Hirschberger}}, \bibinfo {author}
  {\bibfnamefont {E.~V.}\ \bibnamefont {Chulkov}}, \bibinfo {author}
  {\bibfnamefont {A.}~\bibnamefont {Ernst}}, \bibinfo {author} {\bibfnamefont
  {N.~P.}\ \bibnamefont {Ong}}, \bibinfo {author} {\bibfnamefont {R.~J.}\
  \bibnamefont {Cava}}, \ and\ \bibinfo {author} {\bibfnamefont {B.~A.}\
  \bibnamefont {Bernevig}},\ }\href@noop {} {\bibfield  {journal} {\bibinfo
  {journal} {Phys. Rev. Lett.}\ }\textbf {\bibinfo {volume} {117}},\ \bibinfo
  {pages} {236401} (\bibinfo {year} {2016})}\BibitemShut {NoStop}%
\bibitem [{\citenamefont {Jin}\ \emph {et~al.}(2017)\citenamefont {Jin},
  \citenamefont {Wang}, \citenamefont {Chen}, \citenamefont {Zhao},
  \citenamefont {Zhao},\ and\ \citenamefont {Xu}}]{Jin17ferromagnetic}%
  \BibitemOpen
  \bibfield  {author} {\bibinfo {author} {\bibfnamefont {Y.~J.}\ \bibnamefont
  {Jin}}, \bibinfo {author} {\bibfnamefont {R.}~\bibnamefont {Wang}}, \bibinfo
  {author} {\bibfnamefont {Z.~J.}\ \bibnamefont {Chen}}, \bibinfo {author}
  {\bibfnamefont {J.~Z.}\ \bibnamefont {Zhao}}, \bibinfo {author}
  {\bibfnamefont {Y.~J.}\ \bibnamefont {Zhao}}, \ and\ \bibinfo {author}
  {\bibfnamefont {H.}~\bibnamefont {Xu}},\ }\href@noop {} {\bibfield  {journal}
  {\bibinfo  {journal} {Phys. Rev. B}\ }\textbf {\bibinfo {volume} {96}},\
  \bibinfo {pages} {201102(R)} (\bibinfo {year} {2017})}\BibitemShut {NoStop}%
\bibitem [{\citenamefont {Ominato}\ \emph {et~al.}(2017)\citenamefont
  {Ominato}, \citenamefont {Kobayashi},\ and\ \citenamefont
  {Nomura}}]{Ominato17anisotropic}%
  \BibitemOpen
  \bibfield  {author} {\bibinfo {author} {\bibfnamefont {Y.}~\bibnamefont
  {Ominato}}, \bibinfo {author} {\bibfnamefont {K.}~\bibnamefont {Kobayashi}},
  \ and\ \bibinfo {author} {\bibfnamefont {K.}~\bibnamefont {Nomura}},\
  }\href@noop {} {\bibfield  {journal} {\bibinfo  {journal} {Phys. Rev. B}\
  }\textbf {\bibinfo {volume} {95}},\ \bibinfo {pages} {085308} (\bibinfo
  {year} {2017})}\BibitemShut {NoStop}%
\bibitem [{\citenamefont {Kobayashi}\ \emph {et~al.}(2018)\citenamefont
  {Kobayashi}, \citenamefont {Ominato},\ and\ \citenamefont
  {Nomura}}]{Kobayashi18helicity}%
  \BibitemOpen
  \bibfield  {author} {\bibinfo {author} {\bibfnamefont {K.}~\bibnamefont
  {Kobayashi}}, \bibinfo {author} {\bibfnamefont {Y.}~\bibnamefont {Ominato}},
  \ and\ \bibinfo {author} {\bibfnamefont {K.}~\bibnamefont {Nomura}},\
  }\href@noop {} {\bibfield  {journal} {\bibinfo  {journal} {J. Phys. Soc.
  Jpn.}\ }\textbf {\bibinfo {volume} {87}},\ \bibinfo {pages} {073707}
  (\bibinfo {year} {2018})}\BibitemShut {NoStop}%
\bibitem [{\citenamefont {Ye}\ \emph {et~al.}(2018)\citenamefont {Ye},
  \citenamefont {Kang}, \citenamefont {Liu}, \citenamefont {von Cube},
  \citenamefont {Wicker}, \citenamefont {Suzuki}, \citenamefont {Jozwiak},
  \citenamefont {Bostwick}, \citenamefont {Rotenberg}, \citenamefont {Bell},
  \citenamefont {Fu}, \citenamefont {Comin},\ and\ \citenamefont
  {Checkelsky}}]{Ye18massive}%
  \BibitemOpen
  \bibfield  {author} {\bibinfo {author} {\bibfnamefont {L.}~\bibnamefont
  {Ye}}, \bibinfo {author} {\bibfnamefont {M.}~\bibnamefont {Kang}}, \bibinfo
  {author} {\bibfnamefont {J.}~\bibnamefont {Liu}}, \bibinfo {author}
  {\bibfnamefont {F.}~\bibnamefont {von Cube}}, \bibinfo {author}
  {\bibfnamefont {C.~R.}\ \bibnamefont {Wicker}}, \bibinfo {author}
  {\bibfnamefont {T.}~\bibnamefont {Suzuki}}, \bibinfo {author} {\bibfnamefont
  {C.}~\bibnamefont {Jozwiak}}, \bibinfo {author} {\bibfnamefont
  {A.}~\bibnamefont {Bostwick}}, \bibinfo {author} {\bibfnamefont
  {R.}~\bibnamefont {Rotenberg}}, \bibinfo {author} {\bibfnamefont {D.~C.}\
  \bibnamefont {Bell}}, \bibinfo {author} {\bibfnamefont {L.}~\bibnamefont
  {Fu}}, \bibinfo {author} {\bibfnamefont {R.}~\bibnamefont {Comin}}, \ and\
  \bibinfo {author} {\bibfnamefont {J.~G.}\ \bibnamefont {Checkelsky}},\
  }\href@noop {} {\bibfield  {journal} {\bibinfo  {journal} {Nature}\ }\textbf
  {\bibinfo {volume} {555}},\ \bibinfo {pages} {638} (\bibinfo {year}
  {2018})}\BibitemShut {NoStop}%
\bibitem [{\citenamefont {Yin}\ \emph {et~al.}(2018)\citenamefont {Yin},
  \citenamefont {Zhang}, \citenamefont {Li}, \citenamefont {Jiang},
  \citenamefont {Chang}, \citenamefont {Zhang}, \citenamefont {Lian},
  \citenamefont {Xiang}, \citenamefont {Belopolski}, \citenamefont {Zheng},
  \citenamefont {Cochran}, \citenamefont {Xu}, \citenamefont {Bian},
  \citenamefont {Liu}, \citenamefont {Chang}, \citenamefont {Lin},
  \citenamefont {Lu}, \citenamefont {Wang}, \citenamefont {Jia}, \citenamefont
  {Wang},\ and\ \citenamefont {Hasan}}]{Yin18giant}%
  \BibitemOpen
  \bibfield  {author} {\bibinfo {author} {\bibfnamefont {J.-X.}\ \bibnamefont
  {Yin}}, \bibinfo {author} {\bibfnamefont {S.~S.}\ \bibnamefont {Zhang}},
  \bibinfo {author} {\bibfnamefont {H.}~\bibnamefont {Li}}, \bibinfo {author}
  {\bibfnamefont {K.}~\bibnamefont {Jiang}}, \bibinfo {author} {\bibfnamefont
  {G.}~\bibnamefont {Chang}}, \bibinfo {author} {\bibfnamefont
  {B.}~\bibnamefont {Zhang}}, \bibinfo {author} {\bibfnamefont
  {B.}~\bibnamefont {Lian}}, \bibinfo {author} {\bibfnamefont {C.}~\bibnamefont
  {Xiang}}, \bibinfo {author} {\bibfnamefont {I.}~\bibnamefont {Belopolski}},
  \bibinfo {author} {\bibfnamefont {H.}~\bibnamefont {Zheng}}, \bibinfo
  {author} {\bibfnamefont {T.~A.}\ \bibnamefont {Cochran}}, \bibinfo {author}
  {\bibfnamefont {S.-Y.}\ \bibnamefont {Xu}}, \bibinfo {author} {\bibfnamefont
  {G.}~\bibnamefont {Bian}}, \bibinfo {author} {\bibfnamefont {K.}~\bibnamefont
  {Liu}}, \bibinfo {author} {\bibfnamefont {T.-R.}\ \bibnamefont {Chang}},
  \bibinfo {author} {\bibfnamefont {H.}~\bibnamefont {Lin}}, \bibinfo {author}
  {\bibfnamefont {Z.-Y.}\ \bibnamefont {Lu}}, \bibinfo {author} {\bibfnamefont
  {Z.}~\bibnamefont {Wang}}, \bibinfo {author} {\bibfnamefont {S.}~\bibnamefont
  {Jia}}, \bibinfo {author} {\bibfnamefont {W.}~\bibnamefont {Wang}}, \ and\
  \bibinfo {author} {\bibfnamefont {M.~Z.}\ \bibnamefont {Hasan}},\ }\href@noop
  {} {\bibfield  {journal} {\bibinfo  {journal} {Nature}\ }\textbf {\bibinfo
  {volume} {562}},\ \bibinfo {pages} {91} (\bibinfo {year} {2018})}\BibitemShut
  {NoStop}%
\bibitem [{\citenamefont {Liu}\ \emph {et~al.}(2018)\citenamefont {Liu},
  \citenamefont {Sun}, \citenamefont {Kumar}, \citenamefont {Muechler},
  \citenamefont {Sun}, \citenamefont {Jiao}, \citenamefont {Yang},
  \citenamefont {Liu}, \citenamefont {Liang}, \citenamefont {Xu}, \citenamefont
  {Kroder}, \citenamefont {S\"{u}\ss}, \citenamefont {Borrmann}, \citenamefont
  {Shekhar}, \citenamefont {Wang}, \citenamefont {Xi}, \citenamefont {Wang},
  \citenamefont {Schnelle}, \citenamefont {Wirth}, \citenamefont {Chen},
  \citenamefont {Goennenwein},\ and\ \citenamefont {Felser}}]{Liu18giant}%
  \BibitemOpen
  \bibfield  {author} {\bibinfo {author} {\bibfnamefont {E.}~\bibnamefont
  {Liu}}, \bibinfo {author} {\bibfnamefont {Y.}~\bibnamefont {Sun}}, \bibinfo
  {author} {\bibfnamefont {N.}~\bibnamefont {Kumar}}, \bibinfo {author}
  {\bibfnamefont {L.}~\bibnamefont {Muechler}}, \bibinfo {author}
  {\bibfnamefont {A.}~\bibnamefont {Sun}}, \bibinfo {author} {\bibfnamefont
  {L.}~\bibnamefont {Jiao}}, \bibinfo {author} {\bibfnamefont {S.-Y.}\
  \bibnamefont {Yang}}, \bibinfo {author} {\bibfnamefont {D.}~\bibnamefont
  {Liu}}, \bibinfo {author} {\bibfnamefont {A.}~\bibnamefont {Liang}}, \bibinfo
  {author} {\bibfnamefont {Q.}~\bibnamefont {Xu}}, \bibinfo {author}
  {\bibfnamefont {J.}~\bibnamefont {Kroder}}, \bibinfo {author} {\bibfnamefont
  {V.}~\bibnamefont {S\"{u}\ss}}, \bibinfo {author} {\bibfnamefont
  {H.}~\bibnamefont {Borrmann}}, \bibinfo {author} {\bibfnamefont
  {C.}~\bibnamefont {Shekhar}}, \bibinfo {author} {\bibfnamefont
  {Z.}~\bibnamefont {Wang}}, \bibinfo {author} {\bibfnamefont {C.}~\bibnamefont
  {Xi}}, \bibinfo {author} {\bibfnamefont {W.}~\bibnamefont {Wang}}, \bibinfo
  {author} {\bibfnamefont {W.}~\bibnamefont {Schnelle}}, \bibinfo {author}
  {\bibfnamefont {S.}~\bibnamefont {Wirth}}, \bibinfo {author} {\bibfnamefont
  {Y.}~\bibnamefont {Chen}}, \bibinfo {author} {\bibfnamefont {S.~T.~B.}\
  \bibnamefont {Goennenwein}}, \ and\ \bibinfo {author} {\bibfnamefont
  {C.}~\bibnamefont {Felser}},\ }\href@noop {} {\bibfield  {journal} {\bibinfo
  {journal} {Nat. Phys.}\ }\textbf {\bibinfo {volume} {14}},\ \bibinfo {pages}
  {1125} (\bibinfo {year} {2018})}\BibitemShut {NoStop}%
\bibitem [{\citenamefont {Muechler}\ \emph {et~al.}()\citenamefont {Muechler},
  \citenamefont {Liu}, \citenamefont {Xu}, \citenamefont {Felser},\ and\
  \citenamefont {Sun}}]{Muechler18prediction}%
  \BibitemOpen
  \bibfield  {author} {\bibinfo {author} {\bibfnamefont {L.}~\bibnamefont
  {Muechler}}, \bibinfo {author} {\bibfnamefont {E.}~\bibnamefont {Liu}},
  \bibinfo {author} {\bibfnamefont {Q.}~\bibnamefont {Xu}}, \bibinfo {author}
  {\bibfnamefont {C.}~\bibnamefont {Felser}}, \ and\ \bibinfo {author}
  {\bibfnamefont {Y.}~\bibnamefont {Sun}},\ }\href@noop {} {\bibinfo  {journal}
  {arXiv:1712.08115}\ }\BibitemShut {NoStop}%
\bibitem [{\citenamefont {Yin}\ \emph {et~al.}(2019)\citenamefont {Yin},
  \citenamefont {Zhang}, \citenamefont {Chang}, \citenamefont {Wang},
  \citenamefont {Tsirkin}, \citenamefont {Guguchia}, \citenamefont {Lian},
  \citenamefont {Zhou}, \citenamefont {Jiang}, \citenamefont {Belopolski},
  \citenamefont {Shumiya}, \citenamefont {Multer}, \citenamefont {Litskevich},
  \citenamefont {Cochran}, \citenamefont {Lin}, \citenamefont {Wang},
  \citenamefont {Neupert}, \citenamefont {Jia}, \citenamefont {Lei},\ and\
  \citenamefont {Hasan}}]{Yin19negative}%
  \BibitemOpen
\bibfield  {journal} {  }\bibfield  {author} {\bibinfo {author} {\bibfnamefont
  {J.-X.}\ \bibnamefont {Yin}}, \bibinfo {author} {\bibfnamefont {S.~S.}\
  \bibnamefont {Zhang}}, \bibinfo {author} {\bibfnamefont {G.}~\bibnamefont
  {Chang}}, \bibinfo {author} {\bibfnamefont {Q.}~\bibnamefont {Wang}},
  \bibinfo {author} {\bibfnamefont {S.~S.}\ \bibnamefont {Tsirkin}}, \bibinfo
  {author} {\bibfnamefont {Z.}~\bibnamefont {Guguchia}}, \bibinfo {author}
  {\bibfnamefont {B.}~\bibnamefont {Lian}}, \bibinfo {author} {\bibfnamefont
  {H.}~\bibnamefont {Zhou}}, \bibinfo {author} {\bibfnamefont {K.}~\bibnamefont
  {Jiang}}, \bibinfo {author} {\bibfnamefont {I.}~\bibnamefont {Belopolski}},
  \bibinfo {author} {\bibfnamefont {N.}~\bibnamefont {Shumiya}}, \bibinfo
  {author} {\bibfnamefont {D.}~\bibnamefont {Multer}}, \bibinfo {author}
  {\bibfnamefont {M.}~\bibnamefont {Litskevich}}, \bibinfo {author}
  {\bibfnamefont {T.~A.}\ \bibnamefont {Cochran}}, \bibinfo {author}
  {\bibfnamefont {H.}~\bibnamefont {Lin}}, \bibinfo {author} {\bibfnamefont
  {Z.}~\bibnamefont {Wang}}, \bibinfo {author} {\bibfnamefont {T.}~\bibnamefont
  {Neupert}}, \bibinfo {author} {\bibfnamefont {S.}~\bibnamefont {Jia}},
  \bibinfo {author} {\bibfnamefont {H.}~\bibnamefont {Lei}}, \ and\ \bibinfo
  {author} {\bibfnamefont {M.~Z.}\ \bibnamefont {Hasan}},\ }\href@noop {}
  {\bibfield  {journal} {\bibinfo  {journal} {Nat. Phys.}\ }\textbf {\bibinfo
  {volume} {15}},\ \bibinfo {pages} {443} (\bibinfo {year} {2019})}\BibitemShut
  {NoStop}%
\bibitem [{\citenamefont {Ozawa}\ and\ \citenamefont {Nomura}()}]{Ozawa19two}%
  \BibitemOpen
  \bibfield  {author} {\bibinfo {author} {\bibfnamefont {A.}~\bibnamefont
  {Ozawa}}\ and\ \bibinfo {author} {\bibfnamefont {K.}~\bibnamefont {Nomura}},\
  }\href@noop {} {\bibinfo  {journal} {arXiv:1904.08148}\ }\BibitemShut
  {NoStop}%
\bibitem [{\citenamefont {Liu}\ \emph {et~al.}(2019)\citenamefont {Liu},
  \citenamefont {Liang}, \citenamefont {Liu}, \citenamefont {Xu}, \citenamefont
  {Li}, \citenamefont {Chen}, \citenamefont {Pei}, \citenamefont {Shi},
  \citenamefont {Mo}, \citenamefont {Dudin}, \citenamefont {Kim}, \citenamefont
  {Cacho}, \citenamefont {Li}, \citenamefont {Sun}, \citenamefont {Yang},
  \citenamefont {Liu}, \citenamefont {Parkin}, \citenamefont {Felser},\ and\
  \citenamefont {Chen}}]{Liu19magnetic}%
  \BibitemOpen
\bibfield  {journal} {  }\bibfield  {author} {\bibinfo {author} {\bibfnamefont
  {D.~F.}\ \bibnamefont {Liu}}, \bibinfo {author} {\bibfnamefont {A.~J.}\
  \bibnamefont {Liang}}, \bibinfo {author} {\bibfnamefont {E.~K.}\ \bibnamefont
  {Liu}}, \bibinfo {author} {\bibfnamefont {Q.~N.}\ \bibnamefont {Xu}},
  \bibinfo {author} {\bibfnamefont {Y.~W.}\ \bibnamefont {Li}}, \bibinfo
  {author} {\bibfnamefont {C.}~\bibnamefont {Chen}}, \bibinfo {author}
  {\bibfnamefont {D.}~\bibnamefont {Pei}}, \bibinfo {author} {\bibfnamefont
  {W.~J.}\ \bibnamefont {Shi}}, \bibinfo {author} {\bibfnamefont {S.~K.}\
  \bibnamefont {Mo}}, \bibinfo {author} {\bibfnamefont {P.}~\bibnamefont
  {Dudin}}, \bibinfo {author} {\bibfnamefont {T.}~\bibnamefont {Kim}}, \bibinfo
  {author} {\bibfnamefont {C.}~\bibnamefont {Cacho}}, \bibinfo {author}
  {\bibfnamefont {G.}~\bibnamefont {Li}}, \bibinfo {author} {\bibfnamefont
  {Y.}~\bibnamefont {Sun}}, \bibinfo {author} {\bibfnamefont {L.~X.}\
  \bibnamefont {Yang}}, \bibinfo {author} {\bibfnamefont {Z.~K.}\ \bibnamefont
  {Liu}}, \bibinfo {author} {\bibfnamefont {S.~S.~P.}\ \bibnamefont {Parkin}},
  \bibinfo {author} {\bibfnamefont {C.}~\bibnamefont {Felser}}, \ and\ \bibinfo
  {author} {\bibfnamefont {Y.~L.}\ \bibnamefont {Chen}},\ }\href@noop {}
  {\bibfield  {journal} {\bibinfo  {journal} {Science}\ }\textbf {\bibinfo
  {volume} {365}},\ \bibinfo {pages} {1282} (\bibinfo {year}
  {2019})}\BibitemShut {NoStop}%
\bibitem [{\citenamefont {Nakatsuji}\ \emph {et~al.}(2015)\citenamefont
  {Nakatsuji}, \citenamefont {Kiyohara},\ and\ \citenamefont
  {Higo}}]{Nakatsuji15large}%
  \BibitemOpen
  \bibfield  {author} {\bibinfo {author} {\bibfnamefont {S.}~\bibnamefont
  {Nakatsuji}}, \bibinfo {author} {\bibfnamefont {N.}~\bibnamefont {Kiyohara}},
  \ and\ \bibinfo {author} {\bibfnamefont {T.}~\bibnamefont {Higo}},\
  }\href@noop {} {\bibfield  {journal} {\bibinfo  {journal} {Nature}\ }\textbf
  {\bibinfo {volume} {527}},\ \bibinfo {pages} {212} (\bibinfo {year}
  {2015})}\BibitemShut {NoStop}%
\bibitem [{\citenamefont {Yang}\ \emph {et~al.}(2017)\citenamefont {Yang},
  \citenamefont {Sun}, \citenamefont {Zhang}, \citenamefont {Shi},
  \citenamefont {Parkin},\ and\ \citenamefont {Yan}}]{Yang17topological}%
  \BibitemOpen
  \bibfield  {author} {\bibinfo {author} {\bibfnamefont {H.}~\bibnamefont
  {Yang}}, \bibinfo {author} {\bibfnamefont {Y.}~\bibnamefont {Sun}}, \bibinfo
  {author} {\bibfnamefont {Y.}~\bibnamefont {Zhang}}, \bibinfo {author}
  {\bibfnamefont {W.-J.}\ \bibnamefont {Shi}}, \bibinfo {author} {\bibfnamefont
  {S.~S.~P.}\ \bibnamefont {Parkin}}, \ and\ \bibinfo {author} {\bibfnamefont
  {B.}~\bibnamefont {Yan}},\ }\href@noop {} {\bibfield  {journal} {\bibinfo
  {journal} {New J. Phys.}\ }\textbf {\bibinfo {volume} {19}},\ \bibinfo
  {pages} {015008} (\bibinfo {year} {2017})}\BibitemShut {NoStop}%
\bibitem [{\citenamefont {Ito}\ and\ \citenamefont
  {Nomura}(2017)}]{Ito17anomalous}%
  \BibitemOpen
  \bibfield  {author} {\bibinfo {author} {\bibfnamefont {N.}~\bibnamefont
  {Ito}}\ and\ \bibinfo {author} {\bibfnamefont {K.}~\bibnamefont {Nomura}},\
  }\href@noop {} {\bibfield  {journal} {\bibinfo  {journal} {J. Phys. Soc.
  Jpn.}\ }\textbf {\bibinfo {volume} {86}},\ \bibinfo {pages} {063703}
  (\bibinfo {year} {2017})}\BibitemShut {NoStop}%
\bibitem [{\citenamefont {Kane}\ and\ \citenamefont {Mele}(2005)}]{Kane05z2}%
  \BibitemOpen
  \bibfield  {author} {\bibinfo {author} {\bibfnamefont {C.~L.}\ \bibnamefont
  {Kane}}\ and\ \bibinfo {author} {\bibfnamefont {E.~J.}\ \bibnamefont
  {Mele}},\ }\href@noop {} {\bibfield  {journal} {\bibinfo  {journal} {Phys.
  Rev. Lett.}\ }\textbf {\bibinfo {volume} {95}},\ \bibinfo {pages} {146802}
  (\bibinfo {year} {2005})}\BibitemShut {NoStop}%
\bibitem [{\citenamefont {Guo}\ and\ \citenamefont
  {Franz}(2009)}]{Guo09topological}%
  \BibitemOpen
  \bibfield  {author} {\bibinfo {author} {\bibfnamefont {H.-M.}\ \bibnamefont
  {Guo}}\ and\ \bibinfo {author} {\bibfnamefont {M.}~\bibnamefont {Franz}},\
  }\href@noop {} {\bibfield  {journal} {\bibinfo  {journal} {Phys. Rev. B}\
  }\textbf {\bibinfo {volume} {80}},\ \bibinfo {pages} {113102} (\bibinfo
  {year} {2009})}\BibitemShut {NoStop}%
\bibitem [{\citenamefont {Caer}\ \emph {et~al.}(1978)\citenamefont {Caer},
  \citenamefont {Malaman},\ and\ \citenamefont {Roques}}]{Caer78mossbauer}%
  \BibitemOpen
  \bibfield  {author} {\bibinfo {author} {\bibfnamefont {G.~L.}\ \bibnamefont
  {Caer}}, \bibinfo {author} {\bibfnamefont {B.}~\bibnamefont {Malaman}}, \
  and\ \bibinfo {author} {\bibfnamefont {B.}~\bibnamefont {Roques}},\
  }\href@noop {} {\bibfield  {journal} {\bibinfo  {journal} {J. Phys. F: Met.
  Phys.}\ }\textbf {\bibinfo {volume} {8}},\ \bibinfo {pages} {323} (\bibinfo
  {year} {1978})}\BibitemShut {NoStop}%
\bibitem [{\citenamefont {Ando}(1991)}]{Ando91quantum}%
  \BibitemOpen
  \bibfield  {author} {\bibinfo {author} {\bibfnamefont {T.}~\bibnamefont
  {Ando}},\ }\href@noop {} {\bibfield  {journal} {\bibinfo  {journal} {Phys.
  Rev. B}\ }\textbf {\bibinfo {volume} {44}},\ \bibinfo {pages} {8017}
  (\bibinfo {year} {1991})}\BibitemShut {NoStop}%
\bibitem [{\citenamefont {Kandala}\ \emph {et~al.}(2015)\citenamefont
  {Kandala}, \citenamefont {Richardella}, \citenamefont {Richardella},
  \citenamefont {Liu},\ and\ \citenamefont {Samarth}}]{Kandala15giant}%
  \BibitemOpen
  \bibfield  {author} {\bibinfo {author} {\bibfnamefont {A.}~\bibnamefont
  {Kandala}}, \bibinfo {author} {\bibfnamefont {A.}~\bibnamefont
  {Richardella}}, \bibinfo {author} {\bibfnamefont {S.}~\bibnamefont
  {Richardella}}, \bibinfo {author} {\bibfnamefont {C.-X.}\ \bibnamefont
  {Liu}}, \ and\ \bibinfo {author} {\bibfnamefont {N.}~\bibnamefont
  {Samarth}},\ }\href@noop {} {\bibfield  {journal} {\bibinfo  {journal} {Nat.
  Commun.}\ }\textbf {\bibinfo {volume} {6}},\ \bibinfo {pages} {7434}
  (\bibinfo {year} {2015})}\BibitemShut {NoStop}%
\bibitem [{\citenamefont {Kou}\ \emph {et~al.}(2015)\citenamefont {Kou},
  \citenamefont {Pan}, \citenamefont {Wan}, \citenamefont {Fan}, \citenamefont
  {Choi}, \citenamefont {Lee}, \citenamefont {Nie}, \citenamefont {Murata},
  \citenamefont {Shao}, \citenamefont {Zhang},\ and\ \citenamefont
  {Wang}}]{Kou15metal}%
  \BibitemOpen
  \bibfield  {author} {\bibinfo {author} {\bibfnamefont {X.}~\bibnamefont
  {Kou}}, \bibinfo {author} {\bibfnamefont {L.}~\bibnamefont {Pan}}, \bibinfo
  {author} {\bibfnamefont {J.}~\bibnamefont {Wan}}, \bibinfo {author}
  {\bibfnamefont {Y.}~\bibnamefont {Fan}}, \bibinfo {author} {\bibfnamefont
  {E.~S.}\ \bibnamefont {Choi}}, \bibinfo {author} {\bibfnamefont {W.-L.}\
  \bibnamefont {Lee}}, \bibinfo {author} {\bibfnamefont {T.}~\bibnamefont
  {Nie}}, \bibinfo {author} {\bibfnamefont {K.}~\bibnamefont {Murata}},
  \bibinfo {author} {\bibfnamefont {Q.}~\bibnamefont {Shao}}, \bibinfo {author}
  {\bibfnamefont {S.-C.}\ \bibnamefont {Zhang}}, \ and\ \bibinfo {author}
  {\bibfnamefont {K.~L.}\ \bibnamefont {Wang}},\ }\href@noop {} {\bibfield
  {journal} {\bibinfo  {journal} {Nat. Commun.}\ }\textbf {\bibinfo {volume}
  {6}},\ \bibinfo {pages} {8474} (\bibinfo {year} {2015})}\BibitemShut
  {NoStop}%
\bibitem [{\citenamefont {Yasuda}\ \emph {et~al.}(2017)\citenamefont {Yasuda},
  \citenamefont {Mogi}, \citenamefont {Yoshimi}, \citenamefont {Tsukazaki},
  \citenamefont {Takahashi}, \citenamefont {Kawasaki}, \citenamefont {Kagawa},\
  and\ \citenamefont {Tokura}}]{Yasuda17quantized}%
  \BibitemOpen
  \bibfield  {author} {\bibinfo {author} {\bibfnamefont {K.}~\bibnamefont
  {Yasuda}}, \bibinfo {author} {\bibfnamefont {M.}~\bibnamefont {Mogi}},
  \bibinfo {author} {\bibfnamefont {R.}~\bibnamefont {Yoshimi}}, \bibinfo
  {author} {\bibfnamefont {A.}~\bibnamefont {Tsukazaki}}, \bibinfo {author}
  {\bibfnamefont {K.~S.}\ \bibnamefont {Takahashi}}, \bibinfo {author}
  {\bibfnamefont {M.}~\bibnamefont {Kawasaki}}, \bibinfo {author}
  {\bibfnamefont {F.}~\bibnamefont {Kagawa}}, \ and\ \bibinfo {author}
  {\bibfnamefont {Y.}~\bibnamefont {Tokura}},\ }\href@noop {} {\bibfield
  {journal} {\bibinfo  {journal} {Science}\ }\textbf {\bibinfo {volume}
  {358}},\ \bibinfo {pages} {1311} (\bibinfo {year} {2017})}\BibitemShut
  {NoStop}%
\end{thebibliography}%


\end{document}